%Paper: hep-ph/9310359
%From: MURGIA@VAXCA.CA.INFN.IT
%Date: Wed, 27 Oct 1993 12:46:06 +0200 (CET)

%Tex ignores everything on a line after %
%
% Plain TeX dialect;
%
%This version numbers pages at top, first page unnumbered,
%second page number 2, double space
\normalbaselineskip=12pt
\baselineskip=12pt plus 1pt minus 1pt
\magnification=1200
\hsize 15.0truecm \hoffset 0.8 truecm
\vsize 22.5truecm \voffset 0.0 truecm
\input thfont.tex
\nopagenumbers
\headline={\ifnum \pageno=1 \hfil \else\hss\tenrm\folio\hss\fi}
\pageno=1
\overfullrule=0pt

\def\lsim{\mathrel{\rlap{\lower4pt\hbox{\hskip1pt$\sim$}}
    \raise1pt\hbox{$<$}}}         %less than or approx. symbol
\def\gsim{\mathrel{\rlap{\lower4pt\hbox{\hskip1pt$\sim$}}
    \raise1pt\hbox{$>$}}}         %greater than or approx. symbol

%\line{\hfil date} This puts whatever is in parentheses at right
%\line{date \hfil} This puts whatever is in parentheses at left
%\vskip 24pt This puts extra space between lines
%\vskip 12pt This puts extra space between lines, but not so much
%\vfill \eject Use this if you want to begin a new page
%\vfill \eject \end Tex ignores everything after this instruction

\line{\hfil October 1993}
\line{\hfil DFTT 63/93}
\vskip 36pt
\centerline{\bf $\chi_{c2} \to \rho\rho$ decay and the $\rho$
polarization in inclusive processes:}
\centerline{\bf a test of mass effects}
\vskip 36pt
\centerline{Mauro Anselmino}
\vskip 8pt
\centerline{\it Dipartimento di Fisica Teorica, Universit\`a di
Torino}
\centerline{\it and Istituto Nazionale di Fisica Nucleare,
Sezione di Torino}
\centerline{\it Via P. Giuria 1, I--10125 Torino, Italy}
\vskip 20pt
\centerline{Francesco Murgia}
\vskip 8pt
\centerline{\it Istituto Nazionale di Fisica Nucleare, Sezione di
Cagliari}
\centerline{\it Via Ada Negri 18, I--09127 Cagliari, Italy}
\vskip 1.0in
\centerline{\bf ABSTRACT}
\vskip 12pt

We compute the helicity density matrix of $\rho$ vector mesons
produced in the two-body decays of $\chi_{c2}$'s in the framework
of perturbative QCD, allowing for mass corrections. The $\chi_{c2}$'s
are inclusively produced in hadronic interactions via gluon fusion
and are shown to be strongly polarized.
Both the case in which the light quarks inside the final mesons are
treated as massless current quarks and the case in which they are seen
as constituent massive ones are considered. Our results show
significant differences between the two cases; the measurement of
the $\rho$ helicity density matrix should allow a test of the
importance of these mass effects.

\vfill\eject

\normalbaselineskip=18pt
\baselineskip=18pt plus 1pt minus 1pt

\noindent{\bf Introduction }
\vskip12pt
In a previous paper [1] we have studied the polarization of vector
mesons (mainly $\rho$ particles) produced in the exclusive process
$p\bar p \to \chi_{c2} \to MM$. Both steps of the process,
the $p\bar p$ annihilation into $\chi_{c2}$ and the charmonium decay
into vector mesons, can be described in perturbative QCD; we can then
compute the polarization of the $\chi_{c2}$ and the spin density
matrix of the final mesons, $\rho(M)$. The elements of $\rho(M)$
can be measured through the angular distribution of the meson
decay products ({\it e.g.} $\rho \to \pi\pi$) and data can be compared
with theoretical predictions.

As it was stressed in Ref. [1], many non perturbative and higher order
corrections to the usual perturbative QCD scheme are known or expected
to be significant in the $Q^2$ region probed by charmonium decays. In
particular, it was explicitly shown that, for $\chi_{c2} \to \rho\rho$
decays, quark mass corrections are large: the values of $\rho(\rho)$
can discriminate between current quark masses ($m_q \simeq 0$) and
constituent quark masses ($m_q = xm_{\rho}$). The assumption of
constituent rather than current quarks as active partons at the
elementary level is supposed to take into account many non perturbative
effects present in the moderate $Q^2$ region involved.

The experimental determination of $\rho_{\lambda \lambda^{\prime}}
(\rho)$ requires a careful analysis of the angular distributions
of the $\rho \to \pi \pi$ decay, in the $\rho$ rest frame, and it
might be a very difficult task due to the small partial decay widths
$\Gamma(\chi_{c2} \to p\bar p)$ and $\Gamma(\chi_{c2} \to \rho
\rho)$. It would certainly help considering also the inclusive
production of $\chi_{c2}$ in $pp$ interactions, which is more
abundant than the exclusive one in $p\bar p$ processes. One can safely
assume the dominant elementary process underlying $\chi_{c2}$
production in $pp$ scattering to be gluon fusion, $gg \to
\chi_{c2}$; this leads to $\chi_{c2}$ states strongly
polarized, and in a spin state different from that created in $p\bar p$
exclusive annihilations. Again, the spin density matrix of a vector
meson produced in the decay of the polarized $\chi_{c2}$ depends on
whether or not we assume massless quarks; actually, for some elements
of $\rho(\rho)$, the difference between current and constituent quarks
turns out to be bigger in this case than in our previous study [1] and
easier to detect.

Let us then consider the inclusive two-step process
$$h_1h_2 \to \chi_{c2}+X \, ,  \quad\quad \chi_{c2} \to M_1M_2 \,,$$
where $h_1$ and $h_2$ are usually two protons but could actually be any
other hadrons. We recall that the helicity density matrix of the vector
meson is given by
$$ \rho_{\lambda_{_1}\lambda_1^{\prime}}(M_1) =
{1\over N} \sum_{\lambda_2,M,M^{\prime}}
A_{\lambda_1\lambda_2;M}A_{\lambda_1^{\prime}\lambda_{_2};
M^{\prime}}^* \,
\hat \rho_{MM^{\prime}}(\chi_{c2}) \eqno(1) $$
where $N$ is the normalization factor,
$$ N = \sum_{\lambda_1,\lambda_2,M,M^{\prime}}
A_{\lambda_1\lambda_2;M}A_{\lambda_1\lambda_2;M^{\prime}}^* \,
\hat \rho_{MM^{\prime}} \> , \eqno(2) $$
such that Tr$\rho=1$, and the $A_{\lambda_1\lambda_2;M}$'s are the
helicity amplitudes for the $\chi_{c2} \to M_1M_2$ process; their
explicit expressions, taking into account quark masses, can be found
in Ref. [1]. $\hat \rho_{MM^{\prime}}(\chi_{c2})$ is the spin
density matrix of the decaying charmonium state, which we shall now
evaluate.

The production of a charge conjugation even charmonium state, like the
$\chi_{c2}$, in the collision of two high energy
hadrons is mainly due to gluon-gluon fusion, $gg \to \chi_{c2}$;
neglecting the transverse motion of gluons inside the hadrons the
charmonium state is produced along the direction of motion of the two
hadrons, say the $z$-axis. Its helicity can then be only $\pm2$ or 0,
depending on the values of the gluon helicities, $\Lambda_1$ and
$\Lambda_2$. The same is true for the third component $M$ of the spin
of a $\chi_{c2}$ produced at rest in the $gg$ center of mass frame;
the actual explicit expression of the helicity amplitudes for such a
process, $A^{\prime}_{M; \Lambda_1 \Lambda_2}(gg \to \chi_{c2})$
can be found, up to some phases due to time reversal and some colour
factors, in Ref. [2], where the most general expressions of the
helicity amplitudes for the $\chi_{c2} \to \gamma \gamma$ decay
are given. From the usual hard scattering scheme one then obtains [3]
$$\eqalign{
\hat\rho_{0,0} &= {|A^{\prime}_{0;1,1}|^2 \over |A^{\prime}_{0;1,1}|^2
+ |A^{\prime}_{2;1,-1}|^2} \cr
\hat\rho_{2,2} &= \hat\rho_{-2,-2}= {1\over 2} \;
{|A^{\prime}_{2;1,-1}|^2 \over |A^{\prime}_{0;1,1}|^2
+ |A^{\prime}_{2;1,-1}|^2} \cr} \eqno(3)$$
Moreover, from Ref. [2] one can notice that, in the non relativistic
limit for the charmonium wave function, only the
$A^{\prime}_{\pm2; \pm1 \mp1}(gg \to \chi_{c2})$
amplitudes survive. This means that the $\chi_{c2}$
can only be produced, via gluon fusion, in a state with $M=\pm2$,
which entails:
$$\hat\rho_{2,2} = \hat\rho_{-2,-2} = {1\over 2} \eqno(4)$$
with all other elements of $\hat\rho(\chi_{c2})$ being zero. We have
explicitly checked that relativistic corrections, that is non zero
relative momentum of the $c\bar c$ pair inside the charmonium state,
lead to very small values of $\hat\rho_{0,0}$ and insignificant
modifications of Eq. (4).

Our conclusion that the only allowed spin states
for inclusively hadro-produced
$\chi_{c2}$'s are the ones with $M=\pm 2$ relies on the dominance of gluon
fusion as elementary underlying process, which is commonly accepted.
Moreover, one can experimentally check the validity of Eq. (4) by
looking at the angular distribution of the decay process $\chi_{c2}
\to J/\psi \, \gamma$, as it was suggested in Ref. [3] in an attempt
of studying the polarized gluon structure functions inside a polarized
proton. In fact the angular distribution of a $\gamma$ emitted by a
$\chi_{c2}$ state in the above radiative decay is given (assuming,
as usual, an electric dipole transition dominance and integrating over
the azimuthal angle) by:
$$W_{\gamma}(\theta)={\hat\rho_{00}\over 8}(5-3\cos^2\theta) +
{3\hat\rho_{11}\over 8}(3-\cos^2\theta) + {3\hat\rho_{22}\over 4}
(1+\cos^2\theta) \eqno(5)$$
which should easily allow a test of Eq. (4), $\hat\rho_{00}=
\hat\rho_{11}=0, \> \hat\rho_{22}=1/2$.

We also notice that a value of $\hat\rho_{22} = 1/2$ would
help in measuring the polarized gluon distribution functions,
according to the analysis of the authors of Ref. [3], in that it gives
the maximum value of the asymmetry they propose to measure (Eqs. (7) and
(10) of Ref. [3], with $f_+ = 0$).

\vskip 18pt
\goodbreak
\noindent
{\bf Results and comments}
\nobreak
\vskip 12pt
\nobreak
{}From Eqs. (1,2,4) and the explicit expressions of $A_{\lambda_1
\lambda_2;M}$ given in Ref. [1] we can now compute $\rho_{\lambda
\lambda^\prime}(\rho)$. One finds
$$ \eqalignno{ \rho_{1,1} &= \rho_{-1,-1} = \cr
&={1\over 4N}\Bigl\{ |\tilde A_{1,-1}|^2 \, {1\over 4} \,
(1 + 6\cos^2\theta + \cos^4\theta) \; + &(6) \cr
&+|\tilde A_{1,0}|^2 \sin^2\theta \>
(1 + \cos^2\theta) + |\tilde A_{1,1}|^2 \, {3\over 2} \,
\sin^4\theta \Bigr\} \cr
\rho_{1,0} &= \rho_{0,1} =
-\rho_{-1,0} = -\rho_{0,-1} = \cr
&= {1\over 4N}|\tilde A_{1,0}| \, \sin\theta\cos\theta \Bigl\{
-|\tilde A_{1,-1}| \, {1\over 2} \, (3+\cos^2\theta) \; + &(7) \cr
&- \bigl[ |\tilde A_{0,0}|-|\tilde A_{1,1}| \bigr]
\sqrt{{3\over 2}} \sin^2\theta \Bigr\} \cr
\rho_{1,-1} &= \rho_{-1,1} \; = \cr
&= {1\over 4N} \sin^2\theta \Bigl\{
|\tilde A_{1,0}|^2 \sin^2\theta \; + & (8) \cr
&+ \sqrt{{3\over 2}} \, {\rm Re} \bigl[ \tilde A_{1,1}\tilde A^*_{1,-1}
\bigr] (1 + \cos^2\theta) \Bigr\} \cr
\rho_{0,0} &= 1-2\rho_{1,1} & (9) \cr }$$
with
$$ \eqalign{ N &= \bigl[ |\tilde A_{0,0}|^2 + 2|\tilde A_{1,1}|^2
\bigr] \, {3\over 8} \, \sin^4\theta \; \cr
&+ |\tilde A_{1,-1}|^2 \, {1\over 8} \, (1 + 6\cos^2\theta +
\cos^4\theta) + |\tilde A_{1,0}|^2 \sin^2\theta \>
(1 + \cos^2\theta) \cr } \eqno(10) $$
$\theta$ is the $\rho$ production angle with respect to the
$\chi_{c2}$ spin quantization axis ($z$-axis), {\it in the
$\chi_{c2}$ rest frame}.
The $\tilde A_{\lambda,\lambda^\prime}$'s can be found in Ref. [1] and
depend on the meson momentum distribution amplitudes $\varphi(x)$.
In the massless limit the above results simplify to $(\tilde A_{1,1}=
\tilde A_{1,0}= 0)$:
$$ \eqalign{ \rho_{1,1} &= \rho_{-1,-1} =
{1\over 2}{\quad 1\over\displaystyle 1+3{\strut |\tilde A_{0,0}|^2
\over\displaystyle |\tilde A_{1,-1}|^2}
{\strut \sin^4\theta\over\displaystyle 1+6\cos^2\theta+\cos^4\theta} }\cr
\rho_{0,0} &= 1-2\rho_{1,1} \cr
\rho_{\lambda\lambda^\prime} &= 0 \qquad\qquad
\lambda \not= \lambda^\prime \cr } \eqno(11) $$

In order to give numerical estimates we need to specify the expressions
of the distribution amplitudes for longitudinally ($\varphi_L, \>
\lambda = 0$) or transversely ($\varphi_T, \> \lambda = \pm1$)
polarized $\rho$ vector mesons, hidden in the $\tilde
A_{\lambda,\lambda^\prime}$'s. We use here the same choice of two
different sets of such distributions as in Ref. [1], which we repeat
for convenience.
In one case we follow Ref. [4], which gives
$$\eqalignno{
\varphi_L(x) &= 13.2x(1-x)-36x^2(1-x)^2 &(12) \cr
\varphi_T(x) &= 30x^2(1-x)^2 &(13) \cr}$$
with a value of the decay constants [1,4]:
$$ f_L \simeq f_T = 0.2 \; {\rm GeV} \eqno(14) $$

The second choice is the simple symmetric distribution amplitude
$$ \varphi_L(x) = \varphi_T(x) = 6x(1-x) \eqno(15) $$
with, again
$$f_L = f_T = 0.2 \; {\rm GeV} \eqno(16) $$

Our results are shown in Figs. 1-3 where we plot respectively
$\rho_{1,1}$, $\rho_{1,0}$ and $\rho_{1,-1}$ as functions of the
production angle $\theta$. The different curves show the results
corresponding to the two choices of the distribution amplitudes
(Eqs. (12,13) or (15)), both in the case of massless current
quarks and massive constituent ones. There are clear differences
in the different cases, both due to a change in the distribution
amplitudes and in the quark masses; notice, however, that with
massless quarks one has $\rho_{1,0} = \rho_{1,-1} =0$, whereas
massive quarks lead to sizeable values of these quantities,
independently of the distribution amplitudes.

Different values of $\rho_{\lambda \lambda^\prime}$ reflect into
different shapes of the polar and azimuthal angular distributions
of the $\pi$ produced by the $\rho \to \pi\pi$ decay, in the $\rho$
helicity rest frame; actually, these angular distributions are what
is experimentally determined. They are related to the elements of
$\rho(\rho)$ by
$$ \eqalignno{
W(\Theta) &= {3\over 2}\bigl[\rho_{0,0}+
(\rho_{1,1}-\rho_{0,0})\sin^2\Theta \bigr] &(17) \cr
W(\Phi) &= {1\over 2\pi}\bigr[ 1-2\rho_{1,-1}
+4\rho_{1,-1}\sin^2\Phi \bigr] &(18) \cr}$$
where $\Theta$ and $\Phi$ are, respectively, the polar and
azimuthal angles of the $\pi$ as it emerges from the decay of
the $\rho$, in the $\rho$ helicity rest frame. Information on $\rho_{1,0}$
can be obtained from the full angular distribution $W(\Theta,
\Phi)$ [1].

In Figs. 4 and 5 we plot respectively $W(\Theta)$ and $W(\Phi)$,
for $\rho$ mesons produced at an angle $\theta = \pi/2$. Again,
clear differences are visible in the different cases; in particular,
we stress that for massless quarks one would obtain a constant value
of $W(\Phi)$, independently of the choice of distribution amplitudes.
The oscillations shown in Fig. 5, if experimentally observed, would be
a definite argument in favour of non perturbative effects, like
constituent quark masses.

Our conclusions are then similar to those reached in Ref. [1], with
some noticeable advantages: the inclusive production of $\chi_{c2}$ in
$pp$ interactions is certainly more copious and the magnitude of mass
effects is in some cases larger than in exclusive $p\bar p$
annihilations. Experimental investigations and verifications should
then be easier.

\vskip 18pt
\goodbreak
\noindent
{\bf References}
\nobreak
\vskip 12pt
\nobreak

\item{[1]} M.~Anselmino and F.~Murgia, {\it Phys. Rev. D} {\bf 47}
(1993) 3977
\item{[2]} M.~Anselmino, F.~Caruso and S.~Forte, {\it Phys. Rev. D}
{\bf 44} (1991) 1438
\item{[3]} J.L.~Cortes and B.~Pire, {\it Phys. Rev. D} {\bf 38} (1988)
3586
\item{[4]} V.L.~Chernyak and A.R.~Zhitnitsky, {\it Phys. Rep.}
{\bf 112} (1984) 173

\vfill\eject

\noindent{\bf Figure Captions}
\vskip12pt
\item{{\bf Fig. 1}} Values of $\rho_{1,1}(\rho)$ as function
of the $\rho$ meson production angle $\theta$. The different
curves correspond to the following cases:
$m_q \not= 0$ with the symmetric distribution amplitude,
Eqs. (15) (solid curve); $m_q = 0$ with symmetric
distribution amplitude (dot--dashed curve); $m_q \not= 0$
with the Ref. [4] distribution amplitude, Eqs. (12,13)
(dashed curve); $m_q = 0$ with the Ref. [4] distribution
amplitude ( dotted curve).
\vskip6pt
\item{{\bf Fig. 2}} Values of $\rho_{1,0}(\rho)$ as function of
$\theta$; same symbols as in Fig. 1. In case $m_q = 0$, one has
$\rho_{1,0}=0$.
\vskip6pt
\item{{\bf Fig. 3}} Values of $\rho_{1,-1}(\rho)$ as function of
$\theta$; same symbols as in Fig. 1. In case $m_q = 0$, one has
$\rho_{1,-1}=0$.
\vskip6pt
\item{{\bf Fig. 4}} Plot of the angular distribution, $W(\Theta)$,
of the $\pi$ emitted in the $\rho$ decay. The $\rho$ has
been produced at an angle $\theta = 90^\circ$. Same symbols as
in Fig. 1.
\vskip6pt
\item{{\bf Fig. 5}} Plot of the angular distribution, $W(\Phi)$,
of the $\pi$ emitted in the $\rho$ decay. The $\rho$ has
been produced at an angle $\theta = 90^\circ$. Same symbols as
in Fig. 1. In case $m_q=0$, one has $W(\Phi)={1\over 2\pi}$.

\bye